
\magnification 1200
\baselineskip=17pt

\centerline{\bf DEGRADATION OF PHASE COHERENCE BY DEFECTS}
\bigskip
\centerline{\bf IN A TWO-DIMENSIONAL VORTEX LATTICE}
\vskip 50pt
\centerline{J. P. Rodriguez}
\medskip
\centerline{\it Dept. of Physics and Astronomy,
California State University, Los Angeles, CA 90032.}
\vskip 30pt
\centerline  {\bf  Abstract}
\vskip 8pt\noindent
The thermodynamic 
 nature of two-dimensional vortex matter is studied 
through a duality analysis of the $XY$ model over the square
lattice with low uniform frustration.
A phase-coherent vortex lattice state is found at low temperature
if rigid translations are prohibited. 
It shows a non-zero
phase rigidity that is degraded exclusively by the creation
of dislocation pairs.  The unbinding of such pairs
causes the vortex lattice to simultaneously lose phase coherence
and to melt at a continuous (Kosterlitz-Thouless) phase transition.
General phase auto-correlation functions are also computed,
and these results 
 are used to argue for the existence of a continuous
melting transition of vortex matter in layered superconductors.

\bigskip
\noindent
PACS Indices:  74.60.Ec, 61.72.Bd, 74.60.Ge, 61.72.Lk

\vfill\eject

The nature of the mixed phase in layered superconductors
remains  badly
understood theoretically.$^1$
For example, the rich phase diagram shown by high-temperature
superconductors in external magnetic field continues to
yield surprises.$^2$    These systems are layered and extremely
type-II.
The minimum theoretical description of a layered type-II superconductor  
in external magnetic field is one that
neglects both magnetic screening and Josephson coupling,
which is equivalent to a 
stack of isolated two-dimensional (2D) $XY$ models
with uniform frustration.$^{3-5}$
A 2D vortex lattice state is believed to exist at low temperature
in such case.$^1$
Indeed, the elastic medium description 
predicts that the 2D vortex lattice melts continuously into an intermediate
hexatic phase that retains the six-fold rotational symmetry
shown by the solid state.$^{6-8}$  On the contrary,
calculations of the long-range phase correlations in
2D and three-dimensional (3D) vortex lattices that   also
employ the elastic medium approximation  find no evidence
for macroscopic phase coherence due to the appearance of
an infrared divergence.$^{9}$  Controversy therefore
surrounds the theoretical existence   of a vortex lattice state
in pure type-II superconductors.$^{10,11}$

In this Letter, we show that a phase-coherent vortex-lattice state
is in fact possible at low temperature in two dimensions
if rigid translations of the vortex lattice
are prohibited by surface barriers.  
A direct calculation
of the phase correlations in the 2D $XY$ model with low uniform
frustration demonstrates that the infrared divergence  alluded
to above$^{9}$ can be 
removed in a natural and unambiguous way through an
asymptotic regularization scheme.
This result is confirmed independently by a direct calculation
of the phase rigidity.  The Villain approximation
is employed throughout.$^{12 - 14}$  We also demonstrate that phase coherence
is lost at the continuous melting transition as a result of
the unbinding of dislocation pairs and in a manner that 
is parallel to
the loss of the shear rigidity.$^6$   Notably,
the renormalization group associated with the phase rigidity
in the vicinity of the 2D melting transition
is identical to that of 
the shear modulus. 
Last, we show how the difference
between the intermediate hexatic phase and the vortex liquid phase
is reflected by a subtle difference in the phase factor of
the phase auto-correlation functions.

The layered $XY$ model
with low uniform frustration 
provides a qualitatively correct
theoretical description of vortex matter deep inside of
the mixed phase
in  extremely type-II layered superconductors, where
magnetic screening can be neglected.$^{5}$
In the absence of Josephson coupling,
its thermodynamics
is determined by the energy functional
$E_{XY}^{(2)} = - J \sum_{\vec r, \mu}  {\rm cos}
[\phi(\vec r + \hat\mu a) - \phi(\vec r) - A_{\mu}(\vec r)]$
for the superfluid  kinetic energy
of an isolated layer
in terms of the superconducting phase $\phi(\vec r)$,
with local phase rigidity  
$J$
and with vector potential $\vec A = (0, 2 \pi f x / a )$.
Here $f$ is the density of vortices and
 $a$ denotes the square lattice constant.
Monte Carlo simulations of the
above uniformly frustrated $XY$ model indicate that melting of the
low-temperature 2D vortex lattice occurs at a temperature
$k_B T_m \cong J / 20$ for low density of
vortices,$^{3}$  $f$.
Monte Carlo simulations of the
corresponding non-neutral lattice Coulomb gas obtain a
quantitatively similar phase diagram.$^{4}$
This indicates    that the Villain approximation for
the uniformly frustrated $XY$ model in two dimensions,$^{12 - 14}$
which reduces to such a Coulomb gas ensemble, 
is valid at temperatures even as high or higher than the vortex-lattice
melting transition.

The Villain approximation itself for the $XY$ model over the square lattice
amounts to the  expression 
$\langle {\rm exp} [i\sum_{\vec r} p(\vec r) \phi(\vec r)]\rangle =
Z_{\rm V}^{(2)}[p]/Z_{\rm V}^{(2)}[0]$
for any generalized phase auto-correlation function
set by an     integer source field, $p(r)$, 
in terms of the quotient
of the corresponding partition functions
$$Z_{\rm V}^{(2)} [p] = \sum_{\{\vec n (\vec r)\}} \Pi_{\vec r}
\delta[\vec\nabla\cdot\vec n|_{\vec r}
- p (\vec r)]
\cdot
 {\rm exp}\Biggl(-{1\over {2\beta}}
\sum_{\vec r} |\vec n|^2
-i\sum_{\vec r} \vec n
\cdot\vec A \Biggr). \eqno (1)$$
Above, $n_{\mu} (\vec r)$ is an integer field that ranges
over the links, $(\vec r, \mu)$, of the square lattice,
while $\beta = J / k_B T$.  
The regime of validity of  the Villain approximation for the
$XY$ model  is
generally at low temperatures.$^{14}$
The conservation equation
expressed  by the $\delta$-function factors above
can be treated exactly.  After a series of manipulations,
one is led to a factorized expression for general phase auto-correlation
functions in terms of independent spin-wave (sw) and vortex (vx)
contributions:$^{13, 14}$
$$\Bigl\langle  e^{i\sum_{\vec r} p \cdot \phi} \Bigr\rangle =
 C_{\rm sw} [p]
\Bigl\langle e^{
i \sum_{\vec R} (Q - f) \cdot  \Phi_P} \Bigr\rangle_{\rm vx}
e^{-i\int_P \vec A\cdot d\vec l}.
\eqno (2)$$
The spin-wave factor above is equal to an exponential
$C_{\rm sw} [p]  = 
{\rm exp} 
[-\eta_{\rm sw}\sum_{(1,2)} 
p(1) 2\pi G^{(2)}(1,2) p(2)]$
of the Greens function
$G^{(2)} = - \nabla^{-2}$ over the square lattice,  
and it decays algebraicly
with a correlation exponent
$\eta_{\rm sw} = (2\pi\beta)^{-1}$.
Note that$^{13, 14}$
 $2\pi G^{(2)} (1, 2) \cong {\rm ln} (r_0^{\prime}/r_{12})$,
with  $r_0^{\prime} \sim a$. 
The vortex contribution above represents an average
over (integer) vorticity $Q(\vec R)$ on   the dual lattice of
points $\vec R$.  These vortices form a Coulomb gas ensemble with 
a uniform background 
at a  density $f$.
The ensemble is weighted by the Gibbs
distribution set by the Coulomb energy
$$E_{\rm vx} =  2\pi J  \sum_{(a, b)}
[Q(a) - f]\ 2\pi G^{(2)} (a, b)\
[Q(b) - f].\eqno (3)$$
The flux field
$$\Phi_P (\vec R) = - 2\pi
 \sum_{\vec R^{\prime}} G^{(2)}(\vec R - \vec R^{\prime})
\cdot P (\vec R\, ^{\prime}) \eqno (4)$$
that appears in the vortex  average [see  Eq. (2)] is a potential
determined
by the charge distribution
$P (\vec R)$ for an arbitrary  string(s) of 
elementary dipoles
that connect oppositely charged probes, $p  (\vec r) = \pm 1$,
of the correlation function (2)     (see 
refs. 13 and 14) .
Last, the line integral over the vector potential on the right-hand
side of Eq. (2) follows the
path ($P$) traced out by this track of   dipoles.

Before we go on to compute general phase autocorrelations
of the 2D vortex lattice using expression (2),
it is instructive to first compute the phase rigidity.
This quantity is known to be given by the one over the
dielectric constant of the non-neutral Coulomb gas ensemble (3) :$^{4, 15}$
$$\rho_s / J = 1 -  {\rm lim}_{k\rightarrow 0}
(2\pi)^2 \beta
\langle Q_{\vec k} Q_{-\vec k} \rangle /
{\cal N} k^2 a^2. \eqno (5)$$ 
Here 
$Q_{\vec k} = \sum_{\vec R} e^{i\vec k\cdot \vec R} Q (\vec R) $
is the Fourier transform of the vorticity
and ${\cal N}$ denotes the total number of sites on the
square lattice.  
Surface barriers are assumed to prohibit ``floating'' of the
2D vortex lattice.$^{3, 4, 7}$
Now suppose that each vortex is displaced
by $\vec u (\vec R)$ with respect to the zero-temperature
triangular vortex lattice. 
Conservation of vorticity dictates that this displacement
field is related to fluctuations in the density 
of the 2D vortex lattice by
$Q - f  = - \vec\nabla\cdot\vec u$.  
This continuity equation  is understood to be coarse-grain
averaged on a scale larger than the triangular vortex lattice
constant, $a_{\triangle}$.
Incompressibility is the  result: rotational invariance gives
$\langle Q_{\vec k} Q_{-\vec k}\rangle = k^2 
\langle [\sum_{\vec R}^{\prime} \vec u ]^2 \rangle / 2 $
as $k\rightarrow 0$.
Substitution 
into Eq. (5) then yields the formula
$${\rho_s\over{J}} = 1 - {2\pi J\over{k_B T}} \cdot
{\pi \langle [\sum_{\vec R}^{\prime} \vec u ]^2 \rangle\over
{N_{\rm vx} a_{\rm vx}^2}} \eqno (6)$$
for the phase rigidity in terms of fluctuations of the
center of mass of the vortex lattice, where rigid
translations are excluded.
  In obtaining Eq.
(6), we have used 
the trivial identity ${\cal N} a^2 = N_{\rm vx} a_{\rm vx}^2$,
where $N_{\rm vx}$ denotes the total number of vortices, and
where $a_{\rm vx}$ is the square root of the area per
vortex.  Notice that vortex/anti-vortex excitations {\it not}
associated with displacements of the zero-temperature
vortex lattice are neglected, which shall be assumed
throughout.
Monte Carlo simulation results indicate that this 
approximation is valid at melting.$^{4}$
Eq. (6) for the rigidity  can be reduced further by expressing
the displacement field  as a superposition of wave and defect
components:$^{7, 8}$
$\vec u = \vec u_{\rm wv} + \vec u_{\rm df}$.
Waves contribute nothing  to the degradation of phase coherence, 
since $\sum \vec u_{\rm wv} = 0$ if 
rigid translations are excluded. 
We then obtain the final result
$${\rho_s\over{J}} = 1 - {2\pi J\over{k_B T}} \cdot
{\pi \langle [\sum_{\vec R}^{\prime} \vec u_{\rm df}]^2 \rangle\over
{N_{\rm vx} a_{\rm vx}^2}} \eqno (7)$$
for the phase rigidity of a 2D vortex lattice.  Notice that
the degradation of phase coherence is due exclusively to
vortex wandering from defects!  

We now 
resume with the direct 
calculation of the phase auto-correlation functions (2).
The first step to take at this stage is to ``remove'' the
net vorticity of the 2D lattice through a singular gauge
transformation:$^5$
$\langle  e^{i\sum_{\vec r} p \cdot \phi} \rangle =
C_{\rm sw} [p]\cdot C_{\rm vx} [p]
\cdot e^{ i \sum_{\vec R} Q_0 \cdot \Phi_P - i\int_P \vec A\cdot d\vec l}$,
where
$C_{\rm vx} [p] = \langle e^{ i \sum_{\vec R} \delta Q \cdot  \Phi_P}
\rangle_{\rm vx}$ 
represents the vortex component.  
Here $\delta Q = Q - Q_0$ is the fluctuation in the
vorticity  with respect to the
triangular vortex lattice at zero temperature, $Q_0 (\vec R)$. 
It is known that  $C_{\rm vx} [p]$
is independent of the particular shape of the path(s) $P$  connecting
probe points $p$ of the correlation function
because of the neutrality condition
$\sum \delta Q = 0$ (see refs.  13 and 14).
And since both the correlation function (2) and its spin-wave component,
$C_{\rm sw}$, are manifestly path invariant,
the phase factor above that remains after the singular gauge transformation
 must therefore also be path invariant.
We then obtain the formal expression
$$\Bigl\langle  e^{i\sum_{\vec r} p \cdot \phi} \Bigr\rangle =
C_{\rm sw} [p]\cdot C_{\rm vx} [p]
\cdot e^{ i \sum_{\vec r} p\cdot \phi_0},\eqno (8)$$
where the phase     $\phi_0 (\vec r)$ 
should    resemble the zero-temperature
configuration.  The next step is to employ the cummulant expansion,
$C_{\rm vx} [p] = {\rm exp}
[-{1\over 2} \langle ( \sum_{\vec R} \delta Q \cdot  \Phi_P )^2
\rangle_{\rm vx}]$,
for the vortex component, which assumes small
distortions of the 2D vortex lattice.  The latter
suggests the Taylor expansion
$\sum_{\vec R} \delta Q \cdot  \Phi_P \cong
\sum_{\vec R}^{\prime} \vec u\cdot\vec\nabla\Phi_P$ 
for the argument,$^9$
which yields
$\langle ( \sum_{\vec R} \delta Q \cdot  \Phi_P )^2 \rangle_{\rm vx}  =
\sum_{a, b} \vec\nabla\Phi_P|_a \cdot
\langle\vec u (a) \vec u (b)\rangle \cdot
 \vec\nabla \Phi_P|_b$.
The cummulant expansion then yields the  algebraic decay
$C_{\rm vx} [p] = {\rm exp} [  \eta_{\rm vx}
\sum_{(1,2)} p(1){\rm ln} (r_{12} / r_0)\, p(2)]$ 
for the vortex component
in the asymptotic limit
$r_{12}\rightarrow\infty$ and $R_{ab} < \infty$,
with a vortex correlation exponent
$$\eta_{\rm vx} = \pi \Bigl\langle 
\Bigl[\sum_{\vec R}^{\qquad\prime} \vec u \Bigr]^2\Bigr\rangle /
N_{\rm vx} a_{\rm vx}^2,\eqno (9)$$
and with the natural ultraviolet scale
$r_0\sim a_{\rm vx}$.
As in the case of the phase rigidity (7), the separation
of the displacement field into wave and defect components
yields the final formula
$$\eta_{\rm vx} = \pi \Bigl\langle
\Bigl[\sum_{\vec R}^{\qquad\prime} \vec u_{\rm df}\Bigr]^2\Bigr\rangle /
N_{\rm vx} a_{\rm vx}^2,\eqno (10)$$
for the correlation exponent of the vortex component in terms
of vortex wandering due to lattice defects.
Last,    Eq. (8) implies   that phase correlations in the 2D vortex
lattice state decay algebraicly, with a net correlation exponent
equal to $\eta = \eta_{\rm sw} + \eta_{\rm vx}$.
Comparison with the spin-wave result 
$\eta_{\rm sw} =  k_B T / 2\pi J $
suggests the identity
$\eta = k_B T/ 2\pi\rho_s$ between this exponent and
the phase rigidity.$^{16}$  A moment's inspection of Eq. (7)
determines that the identity indeed holds true in the
limit $\eta_{\rm vx}\ll\eta_{\rm sw}$, when defects are
dilute, which is implicit throughout.  
In conclusion, both the present  calculation
of the correlation exponent and the
previous calculation of the phase rigidity are in  complete
agreement.

To compute the fluctuations in the center of mass due to defects [Eq. (10)], 
we shall now take the continuum limit, $f\rightarrow 0$, and
model the 2D vortex lattice (3)  as an incompressible
elastic medium:$^{10, 11}$  $\vec\nabla\cdot\vec u = 0$ and
$E_{\rm vx} = {1\over 2}\nu\int d^2 R\, (\vec\nabla\times\vec u)^2$.
Here $\nu$ represents the local  2D shear modulus.
Neither vacancy/interstitials nor disclinations are important 
at low temperature.$^{7, 8}$
We shall therefore compute first the average 
$\langle |\vec u_{\rm df}|^2 \rangle$
due to the  presence of a {\it single} dislocation pair anywhere
inside of  the 2D vortex lattice.$^6$  Indeed, suppose that a dislocation
pair of extent $\vec R_{12}$ lies a distance $\vec R$ from
the origin.  It is then easy to show that
the displacement there is given asymptotically by
$\vec u_{\rm df} (0) \cong  (\pi R)^{-1} (\vec b\cdot\hat R)
[\hat z\cdot (\hat R\times\vec R_{12})]\,  \hat R$,
where $\pm\vec b$ are the equal and opposite Burger's
vectors of  each dislocation.  
We now quote the elastic interaction energy
due to such a dislocation pair:$^{6}$
$E_{12} = \pi^{-1} \nu b^2 
[{\rm ln} (R_{12} / a_{\rm df}) - {1\over 2} {\rm cos} \, 2 \theta]$,
where $\theta$ denotes the angle between $\vec R_{12}$ and $\vec b$,
and where $a_{\rm df}$ represents the core diameter of a dislocation.
After averaging $|\vec u_{\rm df} (0)|^2$ over the Boltzmann
weight, one obtains the result
$\langle |\vec u_{\rm df}|^2 \rangle \cong
(b^2/4\pi) 
(n_{\rm df} \langle R_{12}^2\rangle - 
{1\over 2} n_{\rm df} \langle R_{12}^2 {\rm cos} \, 2 \theta \rangle )
{\rm ln} (R_0/a_{\rm df})$.
Here  $n_{\rm df}$ denotes the density of dislocations,
while $R_0$ represents an infrared cutoff.$^{17}$    The logarithmic
divergence associated with this scale justifies the
neglect of higher-order multipole corrections to
$\vec u_{\rm df} (0)$.
It also justifies the neglect of the contribution
of   the autocorrelator
$\langle \vec u_{\rm df} \cdot \vec u_{\rm df}^{\, \prime} \rangle$
at different points to fluctuations (10) in the
center of mass.
This is due to the fact ({\it i}) that
$-\langle \vec u_{\rm df} (a) \cdot \vec u_{\rm df} (b)\rangle$
must decay faster than
$[\langle |\vec u_{\rm df}|^2 \rangle/2\pi \, {\rm ln} (  R_0 / a_{\rm df})]
 (a_{\rm vx}/R_{ab})^2$ 
because $\eta_{\rm vx} > 0$, and
to the fact ({\it ii}) that the former autocorrelator is short range	
as a result of disordering by the free dislocation pairs.
Eqs. (7) and (10) then yield 
a degradation of
long-range phase coherence 
$${J\over {\rho_s}} =
 1 +  
{\pi\over 2}  \beta
\Biggl({b^2\over {a_{\rm vx}^2}}\Biggr)
\Biggl (
 n_{\rm df} \langle R_{12}^2\rangle -
 {1\over 2}  n_{\rm df} \langle R_{12}^2 {\rm cos}\, 2 \theta\rangle\Biggr )
{\rm ln} \Biggl({R_0\over {a_{\rm df}}}\Biggr) .\eqno (11)$$
The second term on the right-hand side  above is proportional
to the renormalization
of one over the shear modulus obtained first by 
Kosterlitz and Thouless (KT).$^{6}$
The proportionality constant is set by the estimate
$\nu b^2 \cong (\pi/2 \sqrt 3) J$ for the 
shear modulus of the 2D vortex lattice,$^{10, 11}$ with the Burger's vectors
limited to basis vectors of the triangular vortex lattice
($b = a_{\triangle}$).
We conclude that phase correlations in the 2D vortex lattice 
reflect the {\it same}
continuous 2D melting transition that is monitored by the
macroscopic shear modulus!  
In the limit of a dilute concentration of dislocations,
$n_{\rm df}\rightarrow 0$,
this unique 2D melting transition
occurs at a temperature 
$k_B T_m^{(0)} = \nu b^2 / 4 \pi$, which is about
$J / 14$ using the previous estimate for $\nu$.$^{10, 11}$
Eq. (11) therefore implies that the correlation exponent that characterizes
the algebraic decay of long-range phase coherence    (2) 
acquires a small value 
$\eta_m  = k_B T_m^{(0)} / 2\pi J \cong (28\pi)^{-1}$ 
just below the melting temperature. 

Standard 2D melting theory also predicts  short-range translational order
at temperatures just above $T_m$.$^{6-8}$   
The previous calculation (11) indicates that phase correlations are
also short range at $T > T_m$.
The phase correlation length in the melted state can be obtained
from    the following construction. 
At low temperature, $T < T_m$,
the previous algebraic auto-correlation functions (8)
can be recovered by taking the appropriate
power of a partition function
describing the system of dislocations,
$\vec b (\vec R)$:
$$\Bigl\langle  e^{i\sum_{\vec r} p \cdot \phi} \Bigr\rangle =
g_0^{n_+}
\cdot
(Z_{\rm df} [\vec\eta \vec e] / 
Z_{\rm df} [0])^{J^{\prime}/J}
\cdot
e^{ i \sum_{\vec r} p\cdot \phi_0},\eqno (12) $$
with$^{18}$
$$Z_{\rm df} [\vec \eta \vec e   ] = 
\int {\cal D}\vec \psi \sum_{\{\vec b(\vec R)\}}
{\rm exp} \Bigl[- (k_B T / 2 J^{\prime})
\sum_{\vec R}^{\qquad\prime} (\vec\nabla\vec\psi - \vec\eta \vec e   )^2
+  2\pi i  \sum_{\vec R}^{\qquad\prime} \vec\psi\cdot \vec b / a_{\triangle} 
 \Bigr].  \eqno (13)$$
Above, $\vec \psi (\vec R)$ is a real two-component field that ranges over the
zero-temperature vortex lattice.
The vector field 
$\vec\eta$ represents  tracks  of elementary dipoles
over the triangular vortex lattice
that link    probe points of the correlation function
on the dual honeycomb lattice
(see refs. 13 and 14),
while $\vec e$ represents any unit vector.
The exponent 
$J^{\prime} / J = \nu a_{\triangle}^2 / 2\pi^2 J$
is equal to the ratio of $T_m^{(0)}$ to
$T_c^{(0)} = {\pi\over 2} J / k_B$, which is approximately
equal to $(7\pi)^{-1}$ by the previous estimate for the 2D
shear modulus.  Last, the pre-factor
$g_0\sim 1$ is raised to
the number of positive (or negative) probes, $n_+$,
in the correlation function.
Residual angular interactions between dislocations
are  suppressed by the vector Coulomb gas (13).$^{18}$
It is easy to show that the system of dislocations 
is  in a 
``plasma'' phase
at high temperatures, $T > T_m$.  Following the
standard prescription,$^{13, 18}$ the latter indicates that the
two-point phase auto-correlator (12)  decays exponentially above
the melting temperature, with
a correlation length $\xi_{\rm vx}$ that is 
proportional to the Debye screening
length $\lambda_D$ of the ``plasma''.
The phase correlation length $\xi_{\rm vx}$ therefore diverges
exponentially as the system of vortices begins to freeze
exactly like    
${\rm ln}\, \lambda_D \propto (T - T_m)^{-\bar\nu}$,
with exponent $\bar\nu = 2/5$ in the case (13) when residual
angular interactions are suppressed.$^{18}$

The two-point phase correlator is easily calculated
directly from the Villain model (1) at high temperatures, however.$^{5}$
One obtains
$$\langle e^{i\phi(1)}
e^{-i\phi(2)}\rangle  =
g_0 e^{- r_{12}/\xi_{\rm vx}}
e^{-i\int_1^2 \vec A(\vec r)\cdot d\vec r}\eqno (14)$$
asymptotically for this auto-correlation function 
in the disordered phase, where the phase correlation length
is inversely related to the line-tension
of the strings in the dual theory (1).
Although this result exhibits exponential
decay just like the previous one    (12) at $T > T_m$, it
also exhibits a non-trivial phase factor, unlike the prior
result.  The resolution of this paradox is due to   the existence 
of an  intermediate {\it hexatic} phase just above the
melting point,$^{7}$ which was discovered by
Halperin and Nelson after the initial KT analysis of 2D melting.$^{6}$
The phase correlations (14) in the 
vortex liquid phase at
high temperatures, $T > T_h$, show no vestiges of the zero-temperature
vortex lattice asymptotically.  The intermediate
hexatic phase, $T_m < T < T_h$, retains 
long-range orientational order coincident with the
triangular vortex-lattice state,$^{19}$ and this is reflected
by the trivial phase factor in the phase autocorrelator (8).
These subtleties are summarized by Table I.

We shall close    by reviewing 
arguments for the existence
of a continuous melting transition in the layered 3D $XY$ model
with low uniform frustration, which describes
the mixed phase of layered superconductors.$^{5}$  
The author has recently performed
a partial duality analysis of this system  in the
limit of weak Josephson coupling 
that leads to the partition
function
$Z_{\rm CG} = \sum_{\{ n_z  \}}
(\beta / 2\gamma^{\prime 2})^{N[n_z]} 
\Pi_l C [p_l]$
in terms of an integer link field $n_z (\vec r, l)$
on points $\vec r$ between
adjacent layers  $l$ and $l+1$.  Above, $C [p_l]$ represents the
phase auto-correlation function (2) of an isolated layer $l$
probed at $p_l (\vec r) = n_z (\vec r, l-1) - n_z (\vec r, l)$,
while $N [n_z]$ counts the total
 number of fluxon charges, $n_z = \pm 1$.
The latter system is dilute in the weak-coupling limit
reached at large model  anisotropy parameters,
$\gamma^{\prime}\rightarrow\infty$.
It has been shown in this Letter
that $C[p]$ is formally identical to the corresponding autocorrelator
for the 2D $XY$ model without frustration at criticality
in the vicinity of the 2D continuous melting transition
(see Table I).
Since the layered 3D $XY$ model without frustration
shows a continuous   order/disorder transition,$^{20}$
and since this model is also described by a   partition function
like $Z_{\rm CG}$ at weak coupling,
 we conclude that
the layered 3D $XY$ model with low uniform frustration should 
show an analogous  continuous melting transition at weak coupling.$^{5}$

The author  thanks
Charles Creffield, Angeles Vozmediano, Ed Rezayi
 and Victor Gurarie for discussions.




\vfill\eject
\centerline{\bf References}
\vskip 16 pt

\item {1.} M. Tinkham, Physica C {\bf 235}, 3 (1994).

\item {2.} G.W. Crabtree and D.R. Nelson, Physics Today {\bf 50}, 38
(April 1997).

\item {3.} S.A. Hattel and J.M. Wheatley, Phys. Rev. B {\bf 51}, 11951 (1995).

\item {4.} M. Franz and S. Teitel, Phys. Rev. B {\bf 51}, 6551 (1995).

\item {5.}  J.P. Rodriguez, Phys. Rev. B {\bf 62}, 9117 (2000);
Physica C {\bf 332}, 343 (2000); Europhys. Lett. {\bf 54}, 793 (2001).

\item {6.} J.M. Kosterlitz and D.J. Thouless,
J. Phys. C {\bf 6}, 1181 (1973).

\item {7.} D.R. Nelson and B.I. Halperin, Phys. Rev. B {\bf 19},
2457 (1979).

\item {8.} A.P. Young, Phys. Rev. B {\bf 19}, 1855 (1979).

 
\item {9.} M.A. Moore, Phys. Rev. B
{\bf 45}, 7336 (1992);
G. Baym, Phys. Rev. B {\bf 51}, 11697 (1995).

\item {10.} T. Nattermann and S. Scheidl, Adv. Phys. {\bf 49}, 607 (2000).

\item {11.}  G. Blatter, M.V. Feigel'man, V.B. Geshkenbein, A.I. Larkin,
and V.M. Vinokur, Rev. Mod. Phys. {\bf 66}, 1125 (1994).

\item {12.} J. Villain, J. Physique {\bf 36}, 581 (1975).

\item {13.} J.V. Jos\' e, L.P. Kadanoff, S. Kirkpatrick and
D.R. Nelson, Phys. Rev. B {\bf 16}, 1217 (1977).

\item {14.} C. Itzykson and J.  Drouffe, {\it Statistical Field Theory}, 
vol. 1, (Cambridge Univ.  Press, Cambridge, 1991) chap. 4.

\item {15.}
 P. Minnhagen and G.G. Warren, Phys. Rev. B {\bf 24}, 2526 (1981);
 P. Minnhagen, Rev. Mod. Phys. {\bf 59}, 1001 (1987).

\item {16.} J.M. Kosterlitz, J. Phys. C {\bf 7}, 1046 (1974).

\item {17.}  The logarithmic infrared divergence that appears in Eq. (11) is
benign relative to the algebraic one that $n_{\rm df}^{-1}$ acquires
directly from the Boltzmann weight
(see refs. 6-8).

\item {18.} D. R. Nelson, Phys. Rev. B {\bf 18}, 2318 (1978).

\item {19.}  
Theoretical calculations indicate that an underlying grid can act to
convert the second-order hexatic/liquid phase transition 
into a sharp crossover (see refs. 7 and 8).


\item {20.} S. Hikami and T. Tsuneto, Prog. Theor. Phys. {\bf 63}, 387 (1980);
S.R. Shenoy and B. Chattopadhyay, Phys. Rev. B {\bf 51}, 9129 (1995);
S.W. Pierson, Phys. Rev. B {\bf 54}, 688 (1996).


%

\vfill\eject

\noindent {TABLE I.}  Listed are the phase auto-correlation
functions at long-range and the phase rigidities obtained here for the various
phases of the 2D $XY$ model with a low uniform frustration.
Rigid translations of the vortex lattice are prohibited.

\bigskip\bigskip\bigskip

\vbox{\offinterlineskip
\hrule
\halign{&\vrule#&
  \strut\quad\hfil#\hfil\quad\cr
height2pt&\omit&&\omit&&\omit&&\omit&\cr
&Phase\hfil&&$\langle e^{i\phi(1)}e^{-i\phi(2)}\rangle$\hfil&
&$\rho_s / J$\hfil &&Observations&\cr
height2pt&\omit&&\omit&&\omit&&\omit&\cr
\noalign{\hrule}
height2pt&\omit&&\omit&&\omit&&\omit&\cr
&Vortex Lattice ($T < T_m$)&
&$g_0\, (r_0/r_{12})^{\eta}\, e^{i\phi_0(1)} e^{-i\phi_0(2)}$&
&$\eta_{\rm sw}/\eta$&&solid&\cr
&Hexatic ($T_m < T < T_h$)&
&$g_0\, e^{-r_{12}/\xi_{\rm vx}}\, e^{i\phi_0(1)} e^{-i\phi_0(2)}$&
&$0$&&not rigid&\cr
&Vortex Liquid ($T > T_h$)&
&$g_0\, e^{-r_{12}/\xi_{\rm vx}}\, e^{-i\int_1^2 \vec A\cdot d\vec l}$&
&$0$&&not rigid&\cr
height2pt&\omit&&\omit&&\omit&&\omit&\cr}
\hrule}

\end